\documentclass[reprint,prl,nofootinbib,notitlepage,superscriptaddress]{revtex4-1}
\pdfoutput=1
\usepackage{graphicx}
\usepackage[dvipsnames]{xcolor}
\usepackage{amsmath}
\allowdisplaybreaks
\usepackage{amssymb}

\usepackage{slashed}
\usepackage[utf8]{inputenc}
\usepackage[colorlinks=true,linkcolor=blue,bookmarksopen,bookmarksnumbered]{hyperref}
\usepackage{color}
\usepackage[normalem]{ulem}
\usepackage{soul}
\usepackage{units}
\usepackage{rotating}
\usepackage{hhline,multirow,tabularx}
\usepackage{bm}
\usepackage{hyperref}
\usepackage[export]{adjustbox}
\usepackage{mdframed}
\usepackage{comment}
\usepackage{natbib}
\usepackage{lipsum}

\def\bea#1\eea{\begin{align}#1\end{align}} 

\newcommand{\bef}{\begin{figure}[htb]\centering}
\newcommand{\eef}{\end{figure}}

\newcommand{\nn}{\nonumber}

\def\<{\langle}
\def\>{\rangle}

\def\({\left(}
\def\[{\left[}
\def\){\right)}
\def\]{\right]}

\def\ln{\hbox{ln}}

\newcommand{\qt}{q_{\perp}}

\usepackage{lipsum}
\begin{document}

\title{Three-dimensional imaging in nuclei}
	
\author{Mishary Alrashed}
\email{misharyalrashed@g.ucla.edu}
\affiliation{Department of Physics and Astronomy, University of California, Los Angeles, California 90095, USA}

\author{Daniele Anderle}
\email{dpa@m.scnu.edu.cn}
\affiliation{Guangdong Provincial Key Laboratory of Nuclear Science, Institute of Quantum Matter, South China Normal University, Guangzhou 510006, China}
\affiliation{Guangdong-Hong Kong Joint Laboratory of Quantum Matter,
Southern Nuclear Science Computing Center, South China Normal University, Guangzhou 510006, China}
\affiliation{Department of Physics and Astronomy, University of California, Los Angeles, California 90095, USA}

\author{Zhong-Bo Kang}
\email{zkang@g.ucla.edu}
\affiliation{Department of Physics and Astronomy, University of California, Los Angeles, California 90095, USA}
\affiliation{Mani L. Bhaumik Institute for Theoretical Physics, University of California, Los Angeles, California 90095, USA}
\affiliation{Center for Frontiers in Nuclear Science, Stony Brook University, Stony Brook, New York 11794, USA}

\author{John Terry}
\email{johndterry@physics.ucla.edu}
\affiliation{Department of Physics and Astronomy, University of California, Los Angeles, California 90095, USA}
\affiliation{Mani L. Bhaumik Institute for Theoretical Physics, University of California, Los Angeles, California 90095, USA}

\author{Hongxi Xing}
\email{hxing@m.scnu.edu.cn}
\affiliation{Guangdong Provincial Key Laboratory of Nuclear Science, Institute of Quantum Matter, South China Normal University, Guangzhou 510006, China}
\affiliation{Guangdong-Hong Kong Joint Laboratory of Quantum Matter,
Southern Nuclear Science Computing Center, South China Normal University, Guangzhou 510006, China}

\begin{abstract}
We perform the first simultaneous global QCD extraction of the transverse momentum dependent (TMD) parton distribution functions and the TMD fragmentation functions in nuclei. We have considered the world set of data from semi-inclusive electron-nucleus deep inelastic scattering and Drell-Yan di-lepton production. In total, this data set consists of 126 data points from HERMES, Fermilab, RHIC and LHC. Working at next-to-leading order and next-to-next-to-leading logarithmic accuracy, we achieve a $\chi^2/dof = 1.045$. In this analysis, we quantify the broadening of TMDs in nuclei comparing with those in free nucleons for the first time. We also make predictions for the ongoing JLab 12 GeV program and future EIC measurements.

\end{abstract}
\maketitle
{\it Introduction.} In recent years, quantum 3D imaging of the nucleon has become one of the hottest research topics in nuclear physics~\cite{Accardi:2012qut}. Such information is encoded in the transverse momentum dependent parton distribution functions (TMDPDFs) and significant progress has been made in extracting TMDPDFs for free nucleons from experimental data~\cite{Anselmino:2013lza,Bacchetta:2017gcc,Scimemi:2017etj,Bertone:2019nxa,Scimemi:2019cmh,Bacchetta:2019sam} as a central object in hadronic physics community. On the other hand, the corresponding quantum 3D imaging of a heavy nucleus is still at the primitive stage. Identifying the partonic structure of quarks and gluons in nuclei has remained as one of the most important challenges confronting the nuclear physics community since the pioneering EMC measurements in 1980s~\cite{EuropeanMuon:1983wih}, and has been regarded as one of the major goals in future facilities of electron-ion colliders (EIC)~\cite{Accardi:2012qut,AbdulKhalek:2021gbh,Anderle:2021wcy}. Besides characterizing the non-trivial phenomena of nuclear modification of parton distribution inside bounded nucleons and the associated QCD dynamics, an accurate determination of such initial state nuclear effect is mandatory for providing precise benchmark information in searching for the signal of quark-gluon plasma created in heavy-ion collisions~\cite{Salgado:2016jws}.

Recently, tremendous efforts have been devoted to exploring the one-dimensional collinear nuclear parton distribution functions (nPDFs), see review~\cite{Ethier:2020way}. Due to their non-perturbative nature, nPDFs have to be extracted through global analyses of relevant world data within the collinear factorization formalism~\cite{Collins:1989gx}. Significant progress has been made in this business~\cite{Eskola:1998df,deFlorian:2003qf,Hirai:2007sx,Eskola:2007my,Schienbein:2009kk,AtashbarTehrani:2012xh,Khanpour:2016pph,Eskola:2016oht,Walt:2019slu,Kovarik:2015cma,AbdulKhalek:2019mzd,AbdulKhalek:2020yuc}, and novel nuclear phenomena have been identified, see for instance, EPPS16~\cite{Eskola:2016oht}, nCTEQ~\cite{Kova_k_2016}, nNNPDF~\cite{Khalek_2019}. Although there are theoretical models such as parton branching~\cite{Blanco:2019qbm}, multiple scattering in either intermediate Bjorken-$x$~\cite{Schafer:2013mza,Zhang:2021tcc} or small-$x$ saturation region~\cite{Kovchegov:2015zha}, there remains no effort regarding the global extraction of the nuclear transverse momentum dependent parton distribution functions (nTMDPDFs). 

As demonstrated in both the generalized high-twist factorization formalism~\cite{Liang:2008vz} and the dipole model~\cite{Mueller:2016gko,Mueller:2016xoc}, QCD multiple scattering in the nuclear medium is responsible for the difference between TMDPDFs in bound and free nucleons. This QCD multiple scattering leads to the so-called transverse momentum broadening effect, which manifests itself as the nuclear modification of the transverse momentum of the TMDPDFs in the framework of TMD factorization~\cite{Collins:2011zzd}. As such, while nTMDPDFs represent the 3D partonic imaging of nuclei, they are also crucial for understanding the QCD dynamics of multiple scattering in nuclear medium. The accurate determination of nTMDPDFs is therefore one of the important objectives of the future EICs. Among the major goals of EICs, hadronization in medium is also of particular interest, which has been investigated experimentally such as in HERMES~\cite{Airapetian:2007vu}. Such information is usually described by the nuclear modified fragmentation functions (nFFs) involved in the same collinear factorization as that in vacuum \cite{Sassot:2009sh,Guo:2000nz}. However, how the hadronization is influenced by medium in three-dimensional momentum space, i.e. nuclear modified transverse momentum dependent fragmentation functions (nTMDFFs), has never been explored.

The determination of nTMDPDFs and nTMDFFs (collectively called nTMDs) relies on the corresponding TMD QCD factorization theorem~\cite{Collins:2011zzd} for physical observables that involve two distinct scales, which are required to guarantee both the applicability of pQCD and the sensitivity to the parton's transverse motion. Two well-known observables are the transverse momentum distribution of semi-inclusive hadrons in lepton-nucleus deep inelastic scattering (SIDIS) and of di-lepton in Drell-Yan (DY) processes in proton-nucleus (pA) collisions. These observables have been extensively measured by HERMES~\cite{Airapetian:2007vu}, JLab~\cite{Dudek:2012vr,Burkert:2008rj}, Fermilab~\cite{Alde:1990im, Vasilev:1999fa}, RHIC~\cite{Leung:2018tql} and the LHC~\cite{Khachatryan:2015pzs,Aad:2015gta}, and will be further measured by the future EIC~\cite{Accardi:2012qut,AbdulKhalek:2021gbh,Anderle:2021wcy} with unprecedented precision. 

In this letter, we perform the first simultaneous QCD global analysis for nTMDPDFs and nTMDFFs using the world data from SIDIS and DY processes with nuclei. From our global analysis at next-to-leading order (NLO) and next-to-next-to-leading logarithmic (NNLL) accuracy, we demonstrate for the first time quantitatively the broadening of the transverse momentum of partons in bound nucleons.

{\it TMD factorization formalism.} To perform our global analysis, we select SIDIS and DY processes which have well-established TMD factorization formalism~\cite{Collins:2011zzd}. For $ep$ SIDIS, $e(l)+p(P)\rightarrow e(l')+h(P_h)+X$, the differential cross section at small hadron transverse momentum $P_{h\perp} \ll Q$ can be expressed as follows
\begin{align}
    \label{eq:fac-SIDIS}
    \frac{d\sigma^p}{d\mathcal{PS}} & = \sigma_0^{\rm DIS}\,H^{\rm DIS}(Q,\mu)\, \sum_q e_q^2 \int_0^\infty \frac{b\, db}{2\pi} J_0\left(\frac{b P_{h\perp}}{z}\right) \nn
    \\
    & \hspace{1.5cm} \times f_{q/p}(x,b;\mu,\zeta_1)\, D_{h/q}(z,b;\mu,\zeta_2)\,,
\end{align}
where, as in the standard TMD factorization, the result is written in the coordinate $b$-space that is conjugate to $P_{h\perp}$. We have $d\mathcal{PS} = dx\, dQ^2\, dz\, d^2P_{h\perp}$ with $Q^2 = -(l'-l)^2$, $x$ and $z$ the standard SIDIS kinematic variables, $\sigma_0^{\mathrm{DIS}}$ and $H^{\rm DIS}$ are the Born cross section and the hard function for SIDIS. $f_{q/p}$ is the quark TMDPDF inside a proton while $D_{h/q}$ denotes the TMDFF for $q\to h$, with $\mu$ and $\zeta$ representing the renormalization and rapidity scales for TMDs. For the remainder of this paper, we always take $\mu = \sqrt{\zeta_1}=\sqrt{\zeta_2} = Q$ and replace the explicit dependence in the TMDs by the single scale $Q$. Within the so-called Collins-Soper-Sterman formalism~\cite{Collins:1984kg}, the evolved TMDs take the following form
\begin{align}
\label{eq:TMDPDF}
    f_{q/p}(x,b;Q) = & \left[C_{q\leftarrow i}\otimes f_{i/p}\right] (x, \mu_{b_*})\, e^{-S_{\rm pert} - S_{\rm NP}^f},
    \\
    \label{eq:TMDFF}
    D_{h/q}(z,b;Q) = & \frac{1}{z^2}[\hat{C}_{i\leftarrow q}\otimes D_{h/i}] (z, \mu_{b_*}) e^{-S_{\rm pert} - S_{\rm NP}^D},
\end{align}
where $C_{q\leftarrow i}$ and $\hat{C}_{i\leftarrow q}$ are the Wilson coefficient functions, $\otimes$ denotes the convolution, and $f_{i/p}(x, \mu_{b_*})$ and $D_{h/i}(z, \mu_{b_*})$ are the corresponding collinear PDFs and collinear FFs. Here $\mu_{b_*} = 2e^{-\gamma_E}/b_*$ with $\gamma_E$ the Euler constant represents the natural scale for TMD evolution, while $b_*$ is the standard prescription.

TMD evolution handles the evolution for both the longitudinal momentum fraction ($x,\,z$) and transverse component $P_{h\perp}$ (or $b$ in the coordinate space). The collinear functions in Eqs.~\eqref{eq:TMDPDF} and $\eqref{eq:TMDFF}$ control the $x$ ($z$) evolution via the usual DGLAP evolution equation. On the other hand, the perturbative ($S_{\rm pert}$) and non-perturbative ($S_{\rm NP}^{f,D}$) Sudakov factors depend on $b$ and $Q$, which control the corresponding perturbative (small $b$) and non-perturbative (large $b$) evolution on the parton's transverse momentum component and  eventually resums logarithms in $\ln(Q^2/P_{h\perp}^2)$ after the Fourier transform. While $S_{\rm pert}$ is perturbatively calculable, the non-perturbative Sudakov factors have to be obtained by fitting experimental data and take the following form
\begin{align}
    S_{\rm NP}^{f}(b, Q) & =  g_2(b)\,\ln(\sqrt{Q}/\sqrt{Q_0}) + g_q b^2\,,
\\
    S_{\rm NP}^{D}(z,b, Q) & =  g_2(b)\,\ln(\sqrt{Q}/\sqrt{Q_0}) + g_h {b^2}/{z^2}\,,
\end{align}
where $g_2(b)$ parameterizes the large-$b$ behavior of so-called Collins-Soper evolution kernel and is both universal and independent of the species of external hadrons. We follow~\cite{Echevarria:2020hpy,Kang:2015msa} to set $g_2(b) = g_2\, \ln(b/b_*)$. On the other hand, $g_q$ ($g_h$) represent the intrinsic transverse momentum of the TMDs at the initial scale $Q_0$. In a simple Gaussian model, one typically has $g_q\sim \langle k_\perp^2\rangle/4$~\cite{Aybat:2011zv,Anselmino:2012aa}, likewise for $g_h$. The parameters $g_2$, $g_q$, and $g_h$ in vacuum are all constrained in~\cite{Echevarria:2020hpy} with $Q_0 = \sqrt{2.4}$ GeV.

For the DY process in $pp$ collisions, $p(P_1)+p(P_2)\rightarrow \gamma^*/Z(q)+X$, the cross section in the TMD factorization region can be written as follows
\begin{align}
    \label{eq:fac-DY}
    \frac{d\sigma^p}{d\mathcal{PS}}  = &\, \sigma_0^{\rm DY}\,H^{\rm DY}(Q,\mu)\, \mathcal{P}\left(\eta, p_\perp^{\ell\ell}\right)\, \sum_q c_q(Q) \int_0^\infty \frac{b\, db}{2\pi} \nn
    \\
    & \times J_0\left(b\, q_\perp\right) \, f_{\bar{q}/p}(x_1,b;Q)\, f_{q/p}(x_2,b;Q)\,,
\end{align}
where $d\mathcal{PS} = dQ^2\, dy\, d^2\qt$ with $Q,\,y,\, q_\perp$ the invariant mass, rapidity and transverse momentum of the vector boson, while $c_q(Q)$ denotes the quark coupling to the $\gamma^*/Z$~\cite{Scimemi:2017etj}. The term $\mathcal{P}$ takes into account the kinematic cuts on the transverse momentum, $p_\perp^{\ell\ell}$, and the rapidity, $\eta$, of the final state lepton pair~\cite{Bacchetta:2019sam,Scimemi:2017etj,Scimemi:2019cmh}.

In going from a proton to a nuclear target, we follow the same procedure~\cite{Ethier:2020way} that is used for the nuclear collinear PDFs and FFs and make two assumptions. First, we assume that the TMD factorization takes exactly the same form as in Eq.~\eqref{eq:fac-SIDIS}, except that one replaces the TMDs by the nTMDs. Second, we assume that the perturbative physics for nTMDs and TMDs is the same. Using this assumption, the perturbative TMD evolution which is controlled by $S_{\rm pert}$ would remain intact while the Wilson coefficient functions that enter into the OPE are also unchanged. Correspondingly, we would replace collinear functions in Eqs.~\eqref{eq:TMDPDF} and \eqref{eq:TMDFF} by their nuclear versions. In other words, these collinear functions would be modified at an initial scale $Q=Q_0$ and then evolved to the scale $\mu_{b_*}$ via the same DGLAP evolution. On the other hand, we modify the non-perturbative Sudakov $S_{\rm NP}^{f, D}$ to account for the effects from the nuclear medium. In principle, both $g_2(b)$ and the intrinsic components $g_{q,h}$ could be modified~\cite{Kang:2012am} due to the transverse momentum broadening through the parton multiple scattering in the nuclear target. We assume $g_2(b)$ to be the same as that for the proton target, and only replace $g_{q,h}$ by their corresponding nuclear version $g_{q,h}^A$. The $g_{q,h}^A$ parameters would represent the parton's transverse momentum width inside a nucleus, which in general would depend on nuclear size ($\propto A^{1/3}$), momentum fraction $x$ (or $z$) and the hard scale $Q$~\cite{Kang:2013raa,Ru:2019qvz}, denoted as $g_q^A(x, Q)$ and $g_h^A(z, Q)$. In the small-$x$ or gluon saturation region, they would represent the typical size of saturation scale $Q_s^2$~\cite{Mueller:2016xoc,Gelis:2010nm}.

{\it Global analysis.} Considering the limited data available, in this paper we take the known parameters for collinear nPDFs $f_{i/p}^A(x, Q_0)$ and nFFs $D_{h/i}^A(z, Q_0)$, and perform the fit to extract $g_q^A(x, Q)$ and $g_h^A(z, Q)$. With more data in the future, one can simultaneously fit nuclear collinear functions and transverse modification encoded in $g_{q,h}^A$. Specifically, here we follow the EPPS16~\cite{Eskola:2016oht} parameterization for collinear nPDFs with CT14nlo~\cite{Dulat:2015mca} for the proton PDFs, and we take LIKEn collinear nFFs in Ref.~\cite{Zurita:2021kli} for a nuclear target with the DSS14 parameterization~ \cite{deFlorian:2014xna} for the vacuum FFs. We perform our fit at NLO+NNLL accuracy. In the kinematic region probed by the current data, the $x$ (or $z$) and $Q$-dependence is rather mild, which allows us to use two constant parameters $a_N$ and $b_N$ in the fit:
\begin{align}
\label{eq:nuclear-broad}
g_q^A(x, Q) =\, g_q + a_N\, L\,,
\quad
g_h^A(z, Q) =\, g_h + b_N\, L\,.
\end{align}
where $L = A^{1/3} - 1$. Such a modification is similar to the change of the saturation scale $Q_s^2$ in the nucleus~\cite{Mueller:2016gko,Mueller:2016xoc}.
Thus, within our global analysis below, we have introduced the fit parameters $a_N$ and $b_N$, which characterizes the nuclear broadening for the nTMDs.

For the data, we take SIDIS measurements from HERMES collaboration and DY process from Fermilab, RHIC and the LHC. The HERMES collaboration~\cite{Airapetian:2007vu}, measured the hadron multiplicity ratio $R^A_h = M^A_h/M^D_h$, where the superscript $A$ denotes the species of the nuclear target while $D$ denotes a deuteron. On the other hand, $M^A_h = 2\pi P_{h\perp} \frac{d\sigma^A}{d\mathcal{PS}}/ \frac{d\sigma^A}{dx dQ^2}$, with the numerator given by the nuclear version of Eq.~\eqref{eq:fac-SIDIS}. The denominator is the inclusive DIS cross section, for which we use the APFEL library~\cite{Bertone:2013vaa} at NLO with the collinear nPDFs. The CMS and ATLAS collaborations at the LHC directly measure the transverse momentum distribution for $\gamma^*/Z$ production, and we use the arTeMiDe library~\cite{Scimemi:2017etj} to account for the phase space reduction in $\mathcal{P}$. Finally, for Fermilab and RHIC, experimental measurements were performed for nuclear modification factor $R_{AB} = \frac{d\sigma^A}{d\mathcal{PS}}/\frac{d\sigma^B}{d\mathcal{PS}}$,
with $A$ ($B$) the nuclear mass number of the heavy (lighter) nucleus. 

In order to obtain the numerical values of the parameters $a_N$ and $b_N$, we fit the experimental data using the Minuit package~\cite{1975CoPhC..10..343J}. The normalization factors $\cal{N}$ of the LHC data are accounted for in the definition of the $\chi^2$ according to the procedure of~\cite{deFlorian:2014xna,Eskola:2016oht}.
\begin{figure}
    \centering
    \includegraphics[width = 0.48\textwidth]{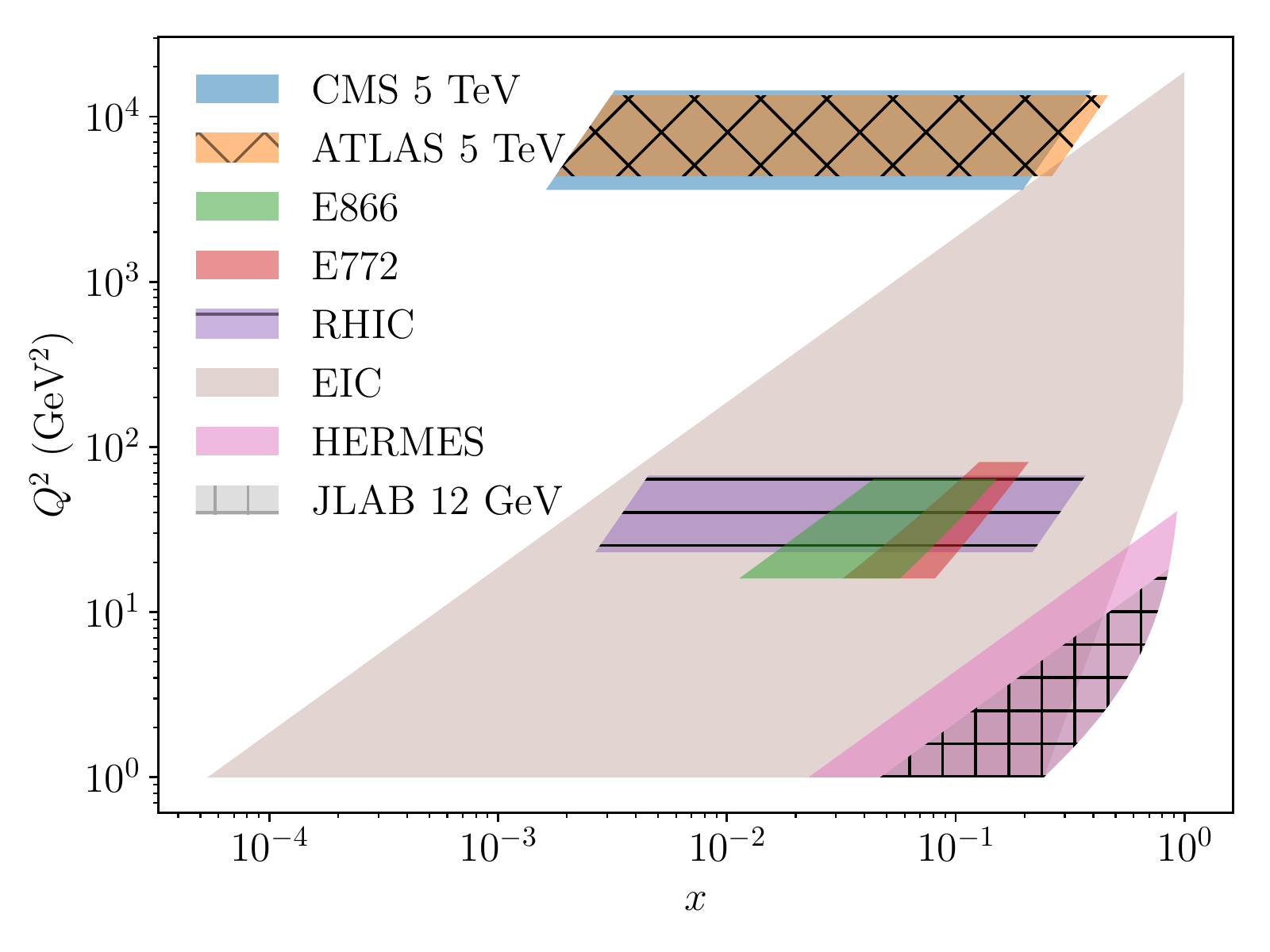}
    \caption{Kinematic coverage for current experimental data and the projected coverage for JLab and the EIC.}
    \label{fig:xvalues}
\end{figure}
In Fig.~\ref{fig:xvalues}, we plot the kinematic coverage of the world data and the kinematic coverage of future experimental data at JLab and the EIC. To select the HERMES data that is within the TMD region, we apply cuts $P_{h \perp}^2 < 0.3\, \textrm{GeV}^2 $ and $ z<0.7$. We note that in order to avoid correlations between the experimental data at HERMES, we choose to fit only one projection of the experimental data. Since the $z$ dependent data provides the largest kinematic coverage, we chose to fit this experimental data. For the DY data, we enforce the standard kinematic cut $q_\perp/Q < 0.3$.
After performing these cuts, we are left with 126 points. 
\begin{figure*}[htb!]
    \centering
    \includegraphics[width = 1 \textwidth]{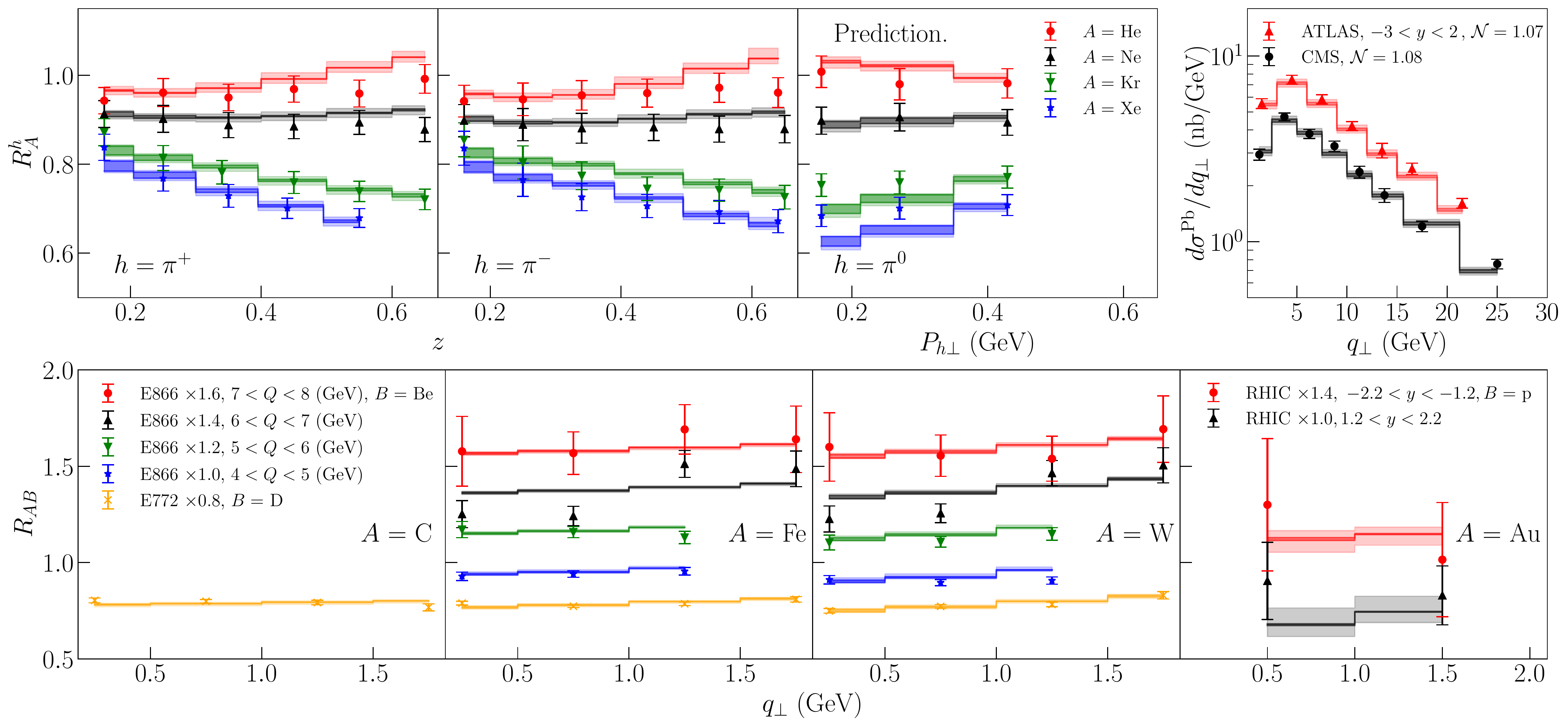}
    \caption{Theoretical description of selected experimental data.}
    \label{fig:data-inc}
\end{figure*} 

{\it Results.} 
\begin{table}[h!]
\centering
 \begin{tabular}{|c c c c c c|} 
 \hline
Collab. & Process & Baseline & Nuclei & $N_{\rm dat}$ & $\chi^2$ \\ [0.5ex] 
 \hline\hline
 HERMES \cite{Airapetian:2007vu} & SIDIS ($\pi$) & D & He,Ne,Kr,Xe & 63 & 40.2 \\ 
 RHIC \cite{Leung:2018tql} & DY & p & Au & 4 & 2.1 \\  
 E772  \cite{Alde:1990im}& DY & D & C,Fe,W & 16 & 22.2  \\ 
 E866  \cite{Vasilev:1999fa} &  DY &  Be & Fe,W & 28 & 42.2  \\ 
 CMS \cite{Khachatryan:2015pzs}  &$\gamma^*/Z$ & p & Pb & 8 & 9.7 \\ 
 ATLAS  \cite{Aad:2015gta} & $\gamma^*/Z$ & p & Pb & 7 & 13.1\\
 {\bf Total} & & & & {\bf 126} & {\bf 129.5}\\
 [1ex] 
 \hline
 \end{tabular}
 \caption{The $\chi^2$ distribution for the data sets used in our fit.}  
 \label{Tab:chi2}
\end{table}
The global analysis of these parameters results in a $\chi^2/d.o.f$ of $1.045$ where the parameter values are given by $a_N = 0.0171 \pm  0.003$ GeV$^2$ and $b_N = 0.0144 \pm 0.001$ GeV$^2$. The $\chi^2$ distribution for the data sets used in our fit are provided in Table~\ref{Tab:chi2}. For the SIDIS data, we study only $\pi$ production. The Baseline column represents the lighter nuclei used in the SIDIS multiplicity ratio and the DY nuclear modification factor. 

In our analysis, we have considered the uncertainty both from the result of our fit to $a_N$ and $b_N$, and the uncertainty from the collinear distributions. In order to generate the fit uncertainties, we use the replica method in Refs.~\cite{Ball:2008by,Signori:2013mda} with 200 replicas. However, at this point we note that we do not consider the collinear uncertainty when generating the replicas. In order to generate the uncertainty from the collinear nPDF and nFF, we use the prescription provided in Ref.~\cite{Eskola:2016oht}. The collinear and fit uncertainties are both displayed at 68$\%$.

In Fig.~\ref{fig:data-inc}, we plot the result of our fit against the experimental data. In the top row of this figure, we plot the comparison against: the multiplicity ratio measurement at HERMES \cite{Airapetian:2007vu} as a function of $z$ (left two columns) and $P_{h\perp}$ (third column from the left), and the DY $q_\perp$ distribution from the LHC (right column). We note that the $P_{h\perp}$ dependent data in the third column is a prediction for those data points. Furthermore, for the LHC data \cite{Khachatryan:2015pzs,Aad:2015gta}, we have provided the $\mathcal{N}_i$ for each of the data sets. In the left three columns of the second row, we plot the comparison against the $R_{AB}$ ratio for the E866~\cite{Vasilev:1999fa} and E772~\cite{Alde:1990im} experiments. Finally, in the right column of this row, we plot the $R_{AB}$ at RHIC~\cite{Leung:2018tql}. In each subplot, we have provided the uncertainty from our fit as a dark band, and the uncertainty from the collinear distributions as a light band.

\begin{figure}[htb!]
    \centering\offinterlineskip
    \includegraphics[height = 0.2\textheight]{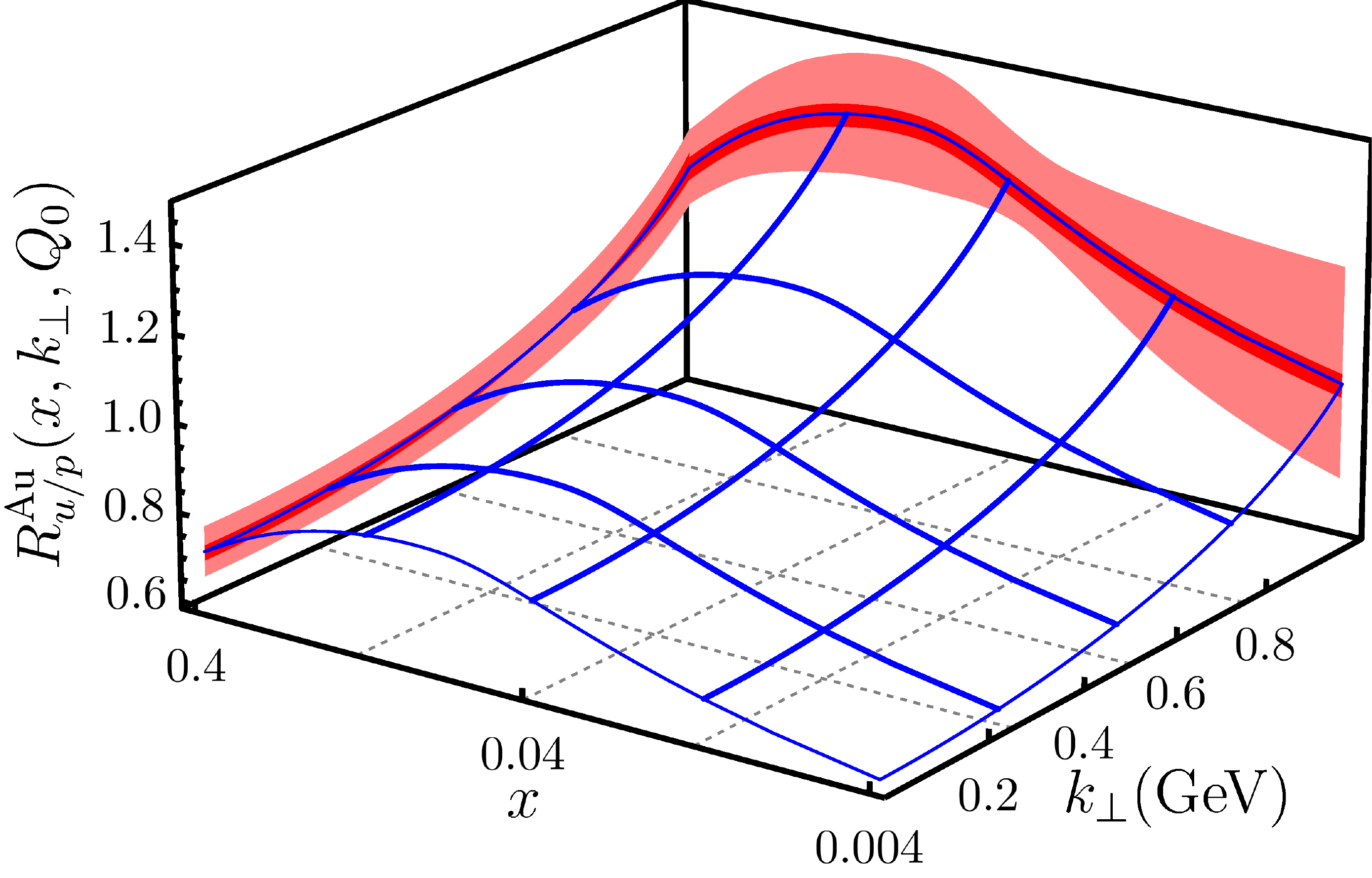}
    \includegraphics[height = 0.2\textheight]{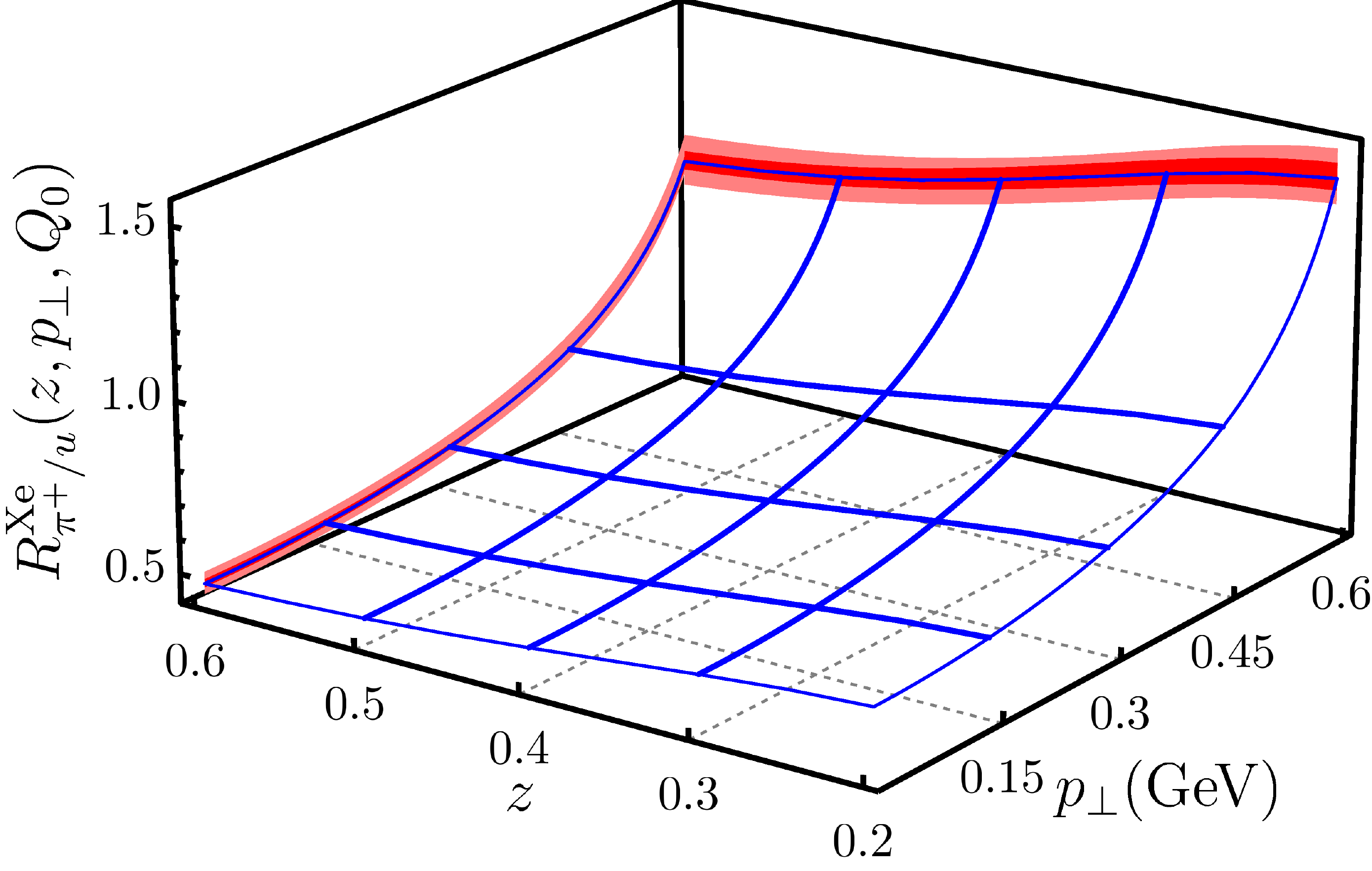}
  \caption{The extracted nuclear ratio for the TMDPDF (top) and the TMDFF (bottom).}
    \label{fig:nTMDs}
\end{figure}

In the top row of Fig.~\ref{fig:nTMDs}, we plot the ratio of the $u$-quark TMDPDF of a bound proton in a gold nucleus and that in a free proton as a function of $x$ and $k_\perp$. The curve along the plane for $k_\perp = 1$ GeV demonstrates the shadowing, anti-shadowing, and the EMC effect which originate from the collinear distribution. The curves which lie in planes of constant $x$ increase with increasing $k_\perp$, which indicates the transverse momentum broadening effect, with a suppression at low $k_{\perp}$ and an enhancement at high $k_{\perp}$, as expected from our model in Eq.~\eqref{eq:nuclear-broad}. In the bottom row of this figure, we plot the ratio of the nTMDFF for $u\rightarrow \pi^+$ in a $\rm Xe$ nucleus and that in vacuum as a function of $z$ and $p_\perp$. Analogous to the nTMDPDFs, we see that as $p_\perp$ grows, this ratio becomes larger, indicating that hadrons originating from fragmentation in the presence of a nuclear medium will tend to have a broader distribution of transverse momentum relative to vacuum TMDFFs. 

\begin{figure}[htb!]
    \centering\offinterlineskip
        \includegraphics[width=0.4\textwidth]{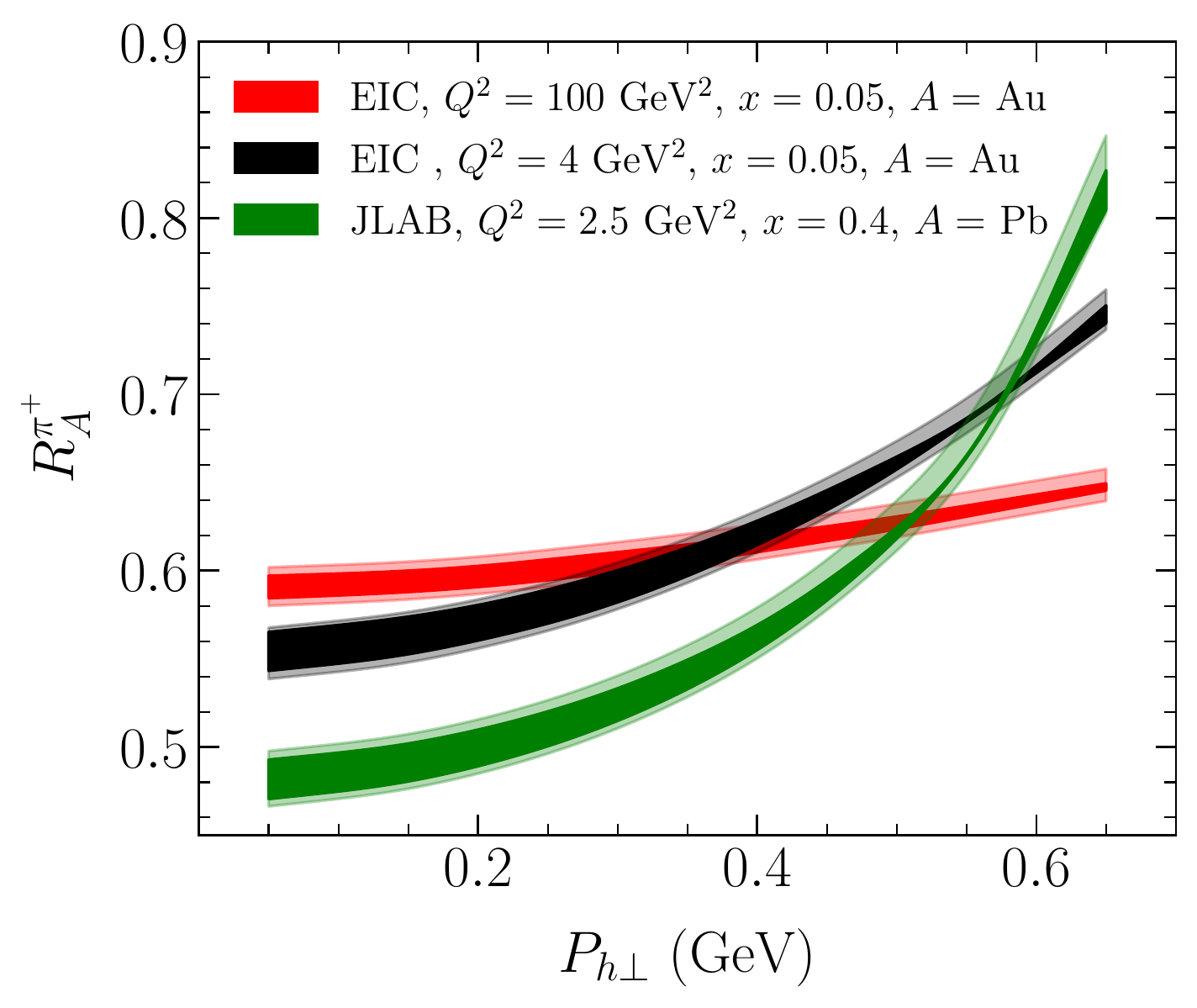}
    \caption{Prediction for the future EIC and JLab at $z =0.4$.}  
    \label{fig:EIC_JLAB}
\end{figure}

In Fig.~\ref{fig:EIC_JLAB}, we plot our prediction for future JLab and EIC multiplicity ratio measurements as a function of $P_{h \perp}$ for $\pi^+$ at $z = 0.4$. For the EIC, we plot the prediction at $x =0.05$ and $Q^2 = 4$ GeV$^2$ (black) and $Q^2 = 100$ GeV$^2$ (red). For JLab, we plot the prediction at $x = 0.4$ and $Q^2 = 2.5$ GeV$^2$ (green). We expect the future measurements will provide a stringent constrain of nTMDs and test the QCD evolution shown in Fig.~\ref{fig:EIC_JLAB}.

{\it Summary.} We perform the first QCD global analysis of nuclear TMDs. For the processes with nuclei, assuming that TMD factorization and perturbative TMD evolution both take the same form as those in the vacuum, at the accuracy of NLO+NNLL we find that we can describe the global set of experimental data using a simple model which accounts for the nonperturbative TMD evolution. We demonstrate quantitatively that both the TMDPDFs and TMDFFs in the presence of the nuclear medium have a broader distribution of transverse momentum. We expect that the framework we have developed will have a large impact on the interpretation of future experimental data at JLab, RHIC, LHC, and the future EICs, allowing us to perform quantum 3D imaging of the nucleus. 

The authors thank Pia Zurita for providing us the LIKEn parameterization. H.X. and D.P.A. are supported by the Guangdong Major Project of Basic and Applied Basic Research No. 2020B0301030008, the Key Project of Science and Technology of Guangzhou (Grant No. 2019050001), the National Natural Science Foundation of China under Grant No. 12022512, No.~12035007. D.P.A. is supported by the China Postdoctoral Science Foundation under Grant No.~2020M672668. M.A. is supported by the UCLA REU program. Z.K. is supported by the National Science Foundation under Grant No.~PHY-1945471. J.T. is supported by NSF Graduate  Research Fellowship Program under Grant No.~DGE-1650604 and UCLA Dissertation Year Fellowship. This work is supported within the framework of the TMD Topical Collaboration.

\bibliography{refs}
\end{document}